\documentclass[12pt]{iopart}

\RequirePackage{mathtools}

\usepackage{iopams}
\usepackage{mathrsfs}
\usepackage{amsfonts}
\usepackage{latexsym}
\usepackage{amsthm}
\usepackage{amsmath}
\usepackage{graphicx}
\usepackage{braket}
\usepackage{xcolor}
\usepackage[colorlinks,linkcolor=blue,citecolor=blue,urlcolor=blue]{hyperref}
\usepackage{subfigure}
\usepackage{enumitem}
\usepackage{scrextend}
\usepackage{multirow}

\begin{document}

\title[]{Quasi-molecular bosonic complexes -- a pathway to atomic analog of SQUID with controlled sensitivity}

\author{
Arghavan Safavi-Naini$^1$, Barbara Capogrosso-Sansone$^{2}$, Anatoly Kuklov$^3$,Vittorio Penna$^4$%\footnote{Corresponding author: fdr@acs.i.kyoto-u.ac.jp}
}
\address{$^1$JILA, National Institute of Standards and Technology and Department of Physics, University of Colorado, 440 UCB, Boulder, CO 80309, USA}
\address{$^2$Department of Physics, Clark University, Worcester, Massachusetts 01610, USA}
\address{$^3$ Department of Engineering \& Physics, CSI, and the Graduate Center of CUNY, New York}
\address{$^4$ Department of Applied Science and Technology and u.d.r. CNISM, Politecnico di Torino, I-10129 Torino, Italy}
\ead{vittorio.penna@polito.it, bcapogrosso@clarku.edu}

\begin{abstract}
Recent experimental advances in realizing degenerate quantum dipolar gases in optical lattices and the flexibility of experimental setups in attaining various geometries offer the opportunity to explore exotic quantum many-body phases stabilized by anisotropic, long-range dipolar interaction. Moreover, the unprecedented control over the various physical properties of these systems, ranging from the quantum statistics of the particles, to the inter-particle interactions, allow one to engineer novel devices. In this paper, we consider dipolar bosons trapped in a stack of one-dimensional optical lattice layers, previously studied in~\cite{Safavi-Naini}. Building on our prior results, we provide a description of the quantum phases stabilized in this system which include composite superfluids, solids, and supercounterfluids, most of which are found to be threshold-less with respect to the dipolar interaction strength. 
We also demonstrate the effect of enhanced sensitivity to rotations of a SQUID-type device made of two composite superfluids  trapped in a ring-shaped optical lattice layer with weak links. 
\end{abstract}

\vspace{2pc}

\maketitle

\section{Introduction}
\label{sec1}
The advances in experimental methods and techniques developed in atomic, molecular and optical physics experiments, such as atomic spectroscopy and interferometry, have led to the development of novel technologies, which have in turn impacted a diverse set of fields, including precision measurement and sensor technologies. Prominent examples include atomic clocks, atomic magnetometers, and atom interferometers~\cite{AtomicSensors}. The recent theoretical and experimental development in atomtronics, allows one to apply techniques from experimental atomic, molecular and optical physics, condensed matter theory, and quantum information to realize ideal analogues to various electronic systems with the atoms playing the role of the electrons. As such, these systems are set to play a significant role in the development of novel quantum technologies and quantum computation~\cite{Holland2007}. 

In the past decade, a variety of atomtronics devices such as atomtronic batteries,
diodes, transistors, atom circuits have been proposed theoretically~\cite{Holland2007,Zoller,Holland2009,Holland2010,Muga,Zozulya,Mompart,Chuanwei,Anderson,Gediminas,Das} and in certain cases realized experimentally~\cite{Hill, Spielman,Esslinger}. More recently, a considerable effort has been devoted to the realization of atomic SQUID (superconducting quantum interference device), which are experimentally created by a Bose-Einstein condensate trapped in a toroidal trap with weak links provided by a variety of techniques such as rotating potential barriers, painted potentials, or Laguerre-Gaussian modes of a laser beam. These systems are analogous to superconducting rings with Josephson junctions, where the response of a superconductor to an external magnetic field is replaced by the response of a superfluid to rotation. 
The neutral versions of the SQUID were first realized in superfluids of $^4$He and $^3$He~\cite{he4,He3}.   
The equivalent of the rf-SQUID has been created in~\cite{CambpellPRL2011,CambpellPRL2013,CambpellNature2014}, while preliminary experiments for a dc-SQUID have been carried out in~\cite{Boshier,CambpellPRL2014}. These experiments have paved the way to the atomtronic rotation sensors analogous to magnetic field sensors formed by superconducting devices. While the atom SQUID experiments listed above were performed in the continuum, recent theoretical proposals call for the realization of atomic rf-SQUIDS in ring-shaped optical lattices~\cite{AmicoSR2014,AmicoNJP2015}. 

Ultracold atoms in optical lattices provide an ideal platform for engineering systems and devices in a highly controllable manner. The unprecedented level of control and flexibility of these experimental setups allow one to construct a variety of system geometries and manipulate inter-particle interactions in an almost ideal, decoherence-free setup. Moreover the experimental realization of atomic and molecular systems featuring long-range and anisotropic dipolar interactions~\cite{Griesmaier2005, Lahaye2007, Ni2008, Ospelkaus2008, Chotia2012, Wu2012, Yan2013, Takekoshi2014, Lu2012, Aikawa2014, Park2015} may lead to the experimental observation of the predicted novel phenomena such as p-wave superfluidity, superfluidity of multimers, solids and supersolids. In what follows we discuss quantum phases of hard-core dipolar bosons trapped in a multi-tube geometry, where each tube is a one-dimensional optical lattice and all tubes are parallel to each other and belong to the same plane. An external field is used to align the dipole moments perpendicular to the tubes. This system features a variety of solid phases, composite superfluids, and supercounterfluids, most of which are predicted to be stabilized at infinitesimal values of dipolar interactions~\cite{Safavi-Naini}. In particular we use bosonization and renormalization group techniques, in association with large scale quantum Monte Carlo simulations, to study and confirm the threshold-less nature of these phases. 

It is important to note that such phases can potentially be used for creating non-linear elements (and their networks) which can be viewed as {\it generalized} Josephson junctions with interesting unexpected features. Unusual properties of a contact between paired and single atomic superfluids were first noticed in Ref.\cite{Kuklov}. 
In general, a contact can be created between arbitrary phases. However, in order to have Josephson effect through such a contact, certain conditions following from the requirement of particle conservation should be met. 

This paper is organized as follows. In Section~\ref{sec2} we describe the microscopic Hamiltonian of the system. Section~\ref{sec3} discusses the quantum phases stabilized by this model. Section~\ref{sec5} focuses the theoretical treatment of the bilayer case, presenting bosonization formalism along with numerical results from quantum Monte Carlo simulations. In Section~\ref{sec4} we summarize results from bosonization and renormalization group analysis for the multilayer system presented in~\cite{Safavi-Naini}. Section~\ref{sec6} is devoted to discussing the nature of Josephson effect between different composite superfluids made of the same type of bosons.
Specifically, we derive a dependence of the rotational "flux" quantization on the number of components forming the composites, and point out that this dependence may be utilized for achieving higher sensitivity in rotational sensors. Finally, in Section~\ref{sec7} we conclude.

\section{Hamiltonian}
\label{sec2}
The system under consideration is described by the model
\begin{equation}
\label{eq:H}
 H= -J \sum_{\langle x,x'\rangle,z}a_{xz}^\dagger a_{x'z}+ \frac{1}{2}\sum_{xz;x'z'} V(x-x',z-z') n_{xz}n_{x'z'}
-\sum_{xz}\mu_{z} \; n_{xz},
\end{equation}
where $J>0$ is the intra-tube tunneling amplitude, $a^\dagger_{xz}$ ($a_{xz}$) is the creation (annihilation) operator for a hard-core boson at site $(x,z)$, where $z=0,1,2, ...,N-1$ labels the tubes, which are parallel to each other and belong to the same plane, and $x=0,1,2,...,L$ is the coordinate along a tube.  The summation over the nearest neighboring sites is denoted by $\sum_{\langle x,x'\rangle,z} ...$. We consider zero inter-tube tunneling, reflected in the absence of terms $a_{xz}^\dagger a_{x'z'}$ with $z'\neq z$, so that the chemical potential $\mu_z$ in layer $z$, controlling its filling factor $\nu_z$, can be chosen to be different in different tubes. Thus, the only coupling between tubes is provided by the dipolar poetntial. Here $n_{xz}= a^\dagger_{xz}a_{xz}$ is particle number operator. Figure~\ref{fig:setup}a) shows a sketch of the system considered.
For the geometry considered here where the polarization axis of the dipoles is  along $z$-axis, the interaction $V(x,z)$ takes the form
\begin{equation}
\label{eq:Vd}
 V(x,z)= V_d \frac{\alpha x^2 - 2z^2}{(\alpha x^2 + z^2)^{5/2}},
\end{equation}
where $V_d>0$ sets the energy scale determined by the inter-tube distance $d_z$ as $V_d \propto 1/d_z^3$, and $\alpha$ denotes the ratio of squares of distances between neighboring sites along $x$ and $z$, respectively. In what follows we use $\alpha=1$. In the case of polarization axis perpendicular to the $xz$-plane, instead, the dipolar interaction is purely repulsive.

Here we consider hard-core bosons (that is, $n_{xz}=0,1$ in Eq.(\ref{eq:H})), which guarantees the stability of the system. For polar molecules trapped within this geometry, the hard-core condition can be satisfied by using a deep enough optical lattice in the $z-$direction, which prevents system collapse due to the attractive forces between aligned dipoles~\cite{Pupillo2007}. For magnetic atoms, one can reach the hard-core limit by exploiting Feshbach resonances which can be used to tune the attractive interactions. Moreover the use of a deep optical lattice along $z$ together with varying $d_z$ can effectively suppress inter-tube tunneling.
\begin{figure}[h]\label{fig:setup}
\begin{center}
\includegraphics[trim={-2cm 7cm 0cm 0.5cm}, clip,width=1.2\textwidth]{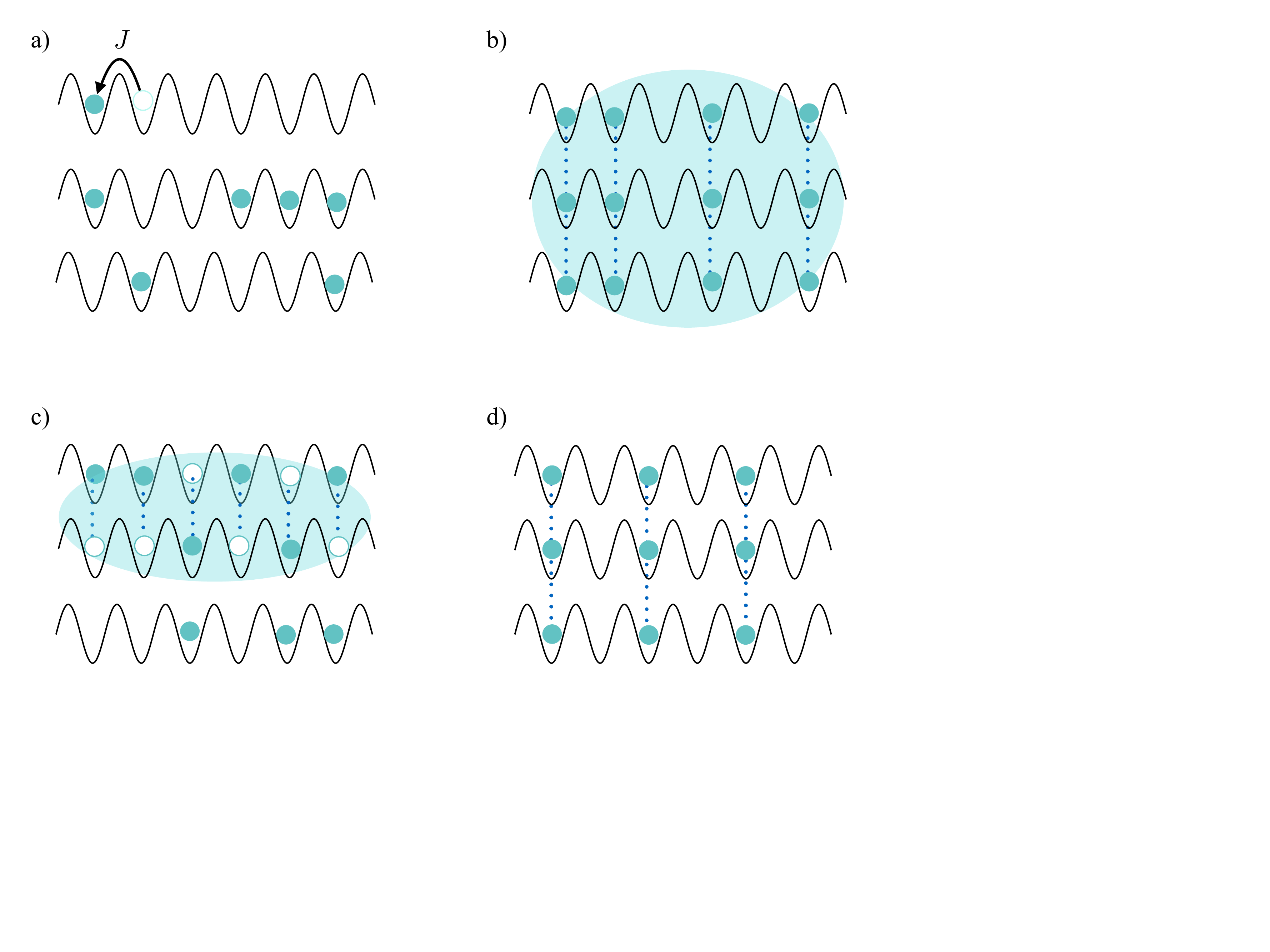}
\caption{Sketches of 1D quantum phases in the multi-tube geometry as described in Sec.\ref{sec3}: a) $N$-SF phase existing for fillings of bosons (solid circles) different  in each tube so that there are independent one-particle algebraic orders in each tube. The solid arrow shows the direction of tunneling along $x$;  b) The CSF phase realized between tubes with equal filling.  Dotted lines indicate binding of bosons from different layers and the fuzzy cloud depicts (three-particle) algebraic order for all layers; c) The SCF phase realized between layers with complementary fillings (upper two) so that the algebraic (two-particle) SCF order (shown by the fuzzy cloud) is formed only on these two layers. Dotted lines indicate binding between particles and holes (open circles). The bottom layer has filling incommensurate with both the upper two layers and, thus, features a one-particle algebraic order;
 d) The CB phase. Dotted lines indicate binding between bosons from different layers at $\nu=1/2$.}
\label{fig:setup}
\end{center}
\end{figure}

In the following we show that in this geometry, where the interaction along the $z$-axis is attractive, any {\em arbitrarily small} $V_d$ can stabilize superfluidity of quasi-molecular complexes, provided that the bosons satisfy the hard-core condition.

\section{Quantum phases in multi-layer geometries} \label{sec3}

In the absence of inter-tube hopping between the layers and the inter-tube interactions, hard-core bosons form $N$ independent superfluids ($N$-SF), which are characterized by $N$ quasi-condensates exhibiting algebraic quantum order.
Within the low energy description the second quantized operators $\psi_z(x)$ built on $a_{xz}$-operators can be replaced 
by phases $\phi_z(x)$ as    $\psi_z(x)  \to  \exp(i \phi_z(x))$, where $\phi_z$ is a variable belonging to the tube $z$. 
In the $N$-SF phase the quantum correlators $\langle \exp(- i\phi_z(x)) \exp(i\phi_{z'}(0))\rangle \propto \delta_{z,z'} |x|^{-b}$,
where $b>0$ is determined by the Luttinger parameter.

As shown in~\cite{Safavi-Naini}, upon introducing an infinitesimal inter-tube interaction, the $N$-SF phase may be destabilized in favor of composite superfluids, supercounterfluids and insulating phases. In a  composite superfluid phase (CSF) the one-particle algebraic order vanishes, that is,  the power law is replaced by the exponential decay $\langle \exp(- i\phi_z(x)) \exp(i\phi_{z'}(0))\rangle \propto \delta_{z,z'} \exp(-|x|/\lambda)$, where $\lambda$ is some length determined by interaction $V_d$.  In the situation of zero threshold $\lambda \to \infty $ as $V_d \to 0$ \cite{Safavi-Naini}. This, in fact, determines the practical experimental limitation -- a system size $L$ should be larger than $\lambda$ in order to observe CSF.  

It is important to note that in the CSF the algebraic order is preserved only  in the corresponding products of the one-particle operators. Here we provide a brief description of the composite quantum many-body phases. 

%As we have shown in~\cite{Safavi-Naini}, due to the hard-core nature of the bosons, an infinitesimal inter-tube coupling stabilizes various superfluid and insulating phases, depending on the filling fractions of the tubes. It should be noted, however, that while these phases are threshold-less, a large spatial scale is required for their observation at arbitrary small $V_d$. In the following we give a brief description of the quantum phases described by model~(\ref{eq:H}).

\subsection{N atomic superfluid ($N$-SF)}\label{AS}

The $N$-SF phase, with $N$ independent atomic superfluids, is stable in the presence of finite $V_d$ if the interacting tubes have fillings $\nu_z$ that are both different and non-complementary. 
%If the filling factor is different between all the tubes and not complementary (to unity)  between interacting tubes, $N$ atomic superfluids will form within the system. 
In this case, there exist $N$ independent one-particle algebraic orders (at $T=0$) in the spatial correlators for each tube as discussed above. This scenario is depicted in Fig.~\ref{fig:setup} a).

\subsection{Composite superfluid (CSF)}\label{CS}
If $n\le N$ tubes in the system have the same incommensurate filling factor, any inter-tube attraction destabilizes the independent atomic superfluids in favor of a superfluid of $n$-boson composites. This phase is characterized by 
an exponential decay of correlators $\langle \exp(i\phi_{z}(x)) \exp(- i\phi_{z}(0))\rangle $ as discussed above, coupled with the survival of the algebraic order in the $n$-body density matrix $\langle \Psi^\dagger(x) \Psi(0)\rangle$ where $\Psi(x) = \psi_1(x)\psi_2(x)...\psi_n(x) $, with $\psi_i,\, i=1,2,...,n$ denoting the fields characterized by the same value of the filling factor $\nu$. This defines the CSF --  a superfluid of quasi-molecular complexes which are formed by $n$ bosons on $n$ tubes. A sketch of this phase is shown in Fig.~\ref{fig:setup}b). It is useful to note that, due to the non-local nature of the interaction, such a phase  can form not  necessarily between adjacent tubes. The only requirement is that the filling factors are the same in a set of $n$ tubes. 

 \subsection{Supercounterfluid (SCF) }\label{SCF}
If $n$ tubes in the system have complementary fillings, the $N$-SF phase can be destabilized in favor of the SCF, previously introduced for a two-component bosonic system in Ref.~\cite{Kuklov2}. SCF can exist in a lattice when the filling factors $\nu_1$ and $\nu_2$ for two components complement each other to an integer filling. In the case of hard-core bosons it is $\nu_1 + \nu_2=1$. Then, the repulsive interaction can induce binding of atoms of the component ``1" to holes of the component ``2". 
This corresponds to the algebraic order in the product $\psi^*_1(x) \psi_2(x)$, while the individual fields $\psi_1$ and $\psi_2$ exhibit no algebraic order.  

This property can be extended to a general case of an $N$-component bosonic mixture where the super-flow of $M<N$ components is (partially) compensated by the counter-flow of the remaining components, provided that the filling factors of the two sets are complementary. The SCF phase is sketched in Fig.~\ref{fig:setup}c) for the case of two (upper) tubes being complementary to each other and the third  one (at the bottom) exhibiting independent one-particle algebraic order. In other words, the SCF algebraic order exists for the fields $\psi^*_3 \psi_2$ and the field $\psi_1$ exhibits one-particle algebraic order, where $z=1,2,3$ from the bottom upward.   The corresponding filling factors obey the relations $\nu_1 \neq \nu_2\neq \nu_3,\, \nu_2+\nu_3=1$.
 
\subsection{Composite insulators}\label{CI} 
The system described by the model~\eqref{eq:H} can support insulating phases at various commensurate fillings
$\nu_z=1/2, 1/3, 2/3, 1/5, ...$. In a single tube, that is, $N=1$, such phases can occur only if $V_d$ exceeds corresponding thresholds. The situation is different for the cases $\nu_z=1/2,\,N>2$.  As shown in Ref.~\cite{Safavi-Naini}, the checker-board (CB) insulator is formed even in the limit   $V_d \to 0$. It is interesting to note that such an insulator is formed even if the inter-layer interaction is purely attractive.   A sketch of the CB solid is shown in Fig.~\ref{fig:setup}d).

\section{Comparison of the bosonization results for $N=2$ tubes with {\it ab initio} QMC }
\label{sec5}

In this Section we review the bosonization approach to the CSF with $N=2$ \cite{Mathey} -- so called paired superfluid (PSF). The bosonization framework \cite{Haldane_1980} is the low energy description of a microscopic model. We  provide detailed comparison with the QMC simulations \cite{Safavi-Naini} using the two-worm algorithm \cite{worm2}. The excellent agreement between the two sets of results provides a strong ground for the extension of the bosonization treatment to arbitrary $N$, as previously presented in ~\cite{Safavi-Naini}. We will explicitly demonstrate that the PSF phase is stabilized by arbitrarily small $V_d$.

\subsection{Renormalization Group (RG) equations for $N$=2 }\label{sec:RGN2}
The PSF in 1D is characterized by algebraic order in the product $\Psi(x)=\psi_1 \psi_2$, 
while the individual fields $\psi_1$ and $\psi_2$ are disordered. We start by reviewing the bosonization approach for a single tube.
The fields $\psi_z$ are represented in terms of the density $\rho_z(x)$ and the conjugated phase $\phi_z$ as $\psi_z(x)=\sqrt{\rho_z(x)}\exp(i\phi_z(x))$, where $x$ is now a continuous variable and   
\begin{equation}
\label{eq:rho}
\rho_z(x)=(\nu_z+\frac{1}{\pi}\nabla_x\theta_z)  \sum_{m=0, \pm 1, ...} e^{2mi(\theta_z+\pi \nu_z x)}\,,
\end{equation}
with $\theta_z(x,t)$ being the  density angle \cite{Haldane_1980}.
In the low energy limit, 1D superfluid can be described by the hydrodynamics of Luttinger liquid. For a single tube (that is, $N=1$)
its classical action in imaginary time $\tau$ can be written as
\begin{equation}
S_1= \frac{1}{2\pi K} \int d\tau dx \left[ \frac{1}{V_s} \dot{\theta}^2_1+ V_s (\partial_x \theta_1)^2\right],
\label{Lutt}
\end{equation}
where $V_s$ stands for speed of sound and $K$ is the Luttinger parameter. It should be mentioned that these parameters are related to details of microscopic theory and finding these relations is not a trivial task. However, for the case of hard-core bosons these relations are known. In particular, in the absence of interaction (except for the hard-core condition), $K=1$ and $V_s=J$ (in the atomic units) (see in \cite{Giamarchi_book}). As we will discuss below the relation $K=1$ turns out to be crucial for the threshold-less nature of some of the composite phases. This important aspect has already been emphasized in Ref.\cite{Mathey} for the case of PSF in $N=2$ system. 

In presence of a periodic lattice and interactions between bosons (in addition to the hard-core one), the higher harmonics in Eq.(\ref{eq:rho}) must be taken into account. Then, the  density-density interaction $V$ (on different sites)  introduces a variety of non-linear terms $ \sim  V \cos(2m'(\theta_1 + \pi \nu_1 x) ) \cos(2m(\theta_1 + \pi \nu_1 x))$  inducing the so called backscattering (see in Ref.~\cite{Giamarchi_book}), with $V$ being the backscattering amplitudes determined by interaction (as $V\propto V_d$ in our case). These terms may become relevant if $\nu_1$ is commensurate with lattice period and $K$ is below some critical value $K_c$. For $\nu_1=1/2$ the lowest relevant harmonic is the $m=\pm 1$ with the corresponding critical value $K_c=1/2$. Thus, a weak interaction $V$ between hard-core bosons does not destroy 1D superfluidity in a single tube because $K=1 > 1/2$. However, as the interaction becomes stronger, $K$ decreases and eventually the insulating CB phase appears. As discussed in Ref.~\cite{Safavi-Naini}, the situation is dramatically different for the case $N>2$ since the critical value of $K$ is unity, $K_c=1$. Hence for $N>2$ an arbitrarily small dipole interaction between tubes induces the CB phase.   

Here, we consider the case $N=2$ and $\nu_1=\nu_2 \neq 1/2$, previously studied in detail by bosonization in Ref.~\cite{Mathey}. We test these predictions by means of QMC simulations and find perfect agreement. For the sake of consistency we repeat the derivations so that the ``language" of the comparison with QMC becomes clear.

The low energy action $S$ for $N=2$ tubes with equal fillings $\nu_{1,2}=\nu$ corresponding to the Hamiltonian (\ref{eq:H}) and which determines the partition function $\sim \exp(-S)$ can be expressed in terms of the variables $\theta_1$ and $\theta_2$ as  
 $$S= S_+ + S_-,$$ 
where the inter-tube interaction $\sim V \rho_1\rho_2 $ is due to the dipolar interaction. The relationship between $V$ and the interaction $V_d$ given by Eq.~\eqref{eq:Vd} is, in general, non-trivial. For weak interactions, the low energy limit is sensitive to the lowest harmonic of interaction for the so called {\it forward} scattering interaction $V_{fs}$ (to be used for the $m=0$ harmonic in Eq.\eqref{eq:rho}) as well as to the backscattering amplitude $V_{bs}$~\cite{Safavi-Naini}. In the limit $V_d\to 0$ it is natural to expect that both $V_{fs}$ and $V_{bs}$ are $ \propto V_d$. In particular, the inter-tube forward scattering interaction term takes the form
$$S_{12}=  \int d\tau dx V_{fs}  \partial_x \theta_1 \partial_x \theta_2,\,\, V_{fs}=\mp \frac{V_d}{\pi^2}$$
where ``-" is for the case of attractive interaction between the tubes (that is, the dipoles are along $z$-axis, see Eq.~\ref{eq:Vd}) and ``+" for the repulsive  interaction realized when the dipoles are perpendicular to the  $xz$-plane. This terms should be added to Eq.(\ref{Lutt}) for both tubes. In what follows we consider the case of attractive interaction, that is, $V_{fs} <0$.

This additional term $S_{12}$ changes the Luttinger parameters for channels $\theta_+=\theta_1 + \theta_2$ and $\theta_-=\theta_1 - \theta_2$ as
\begin{equation}
\label{eq:Kpm}
K_\pm=\frac{2K}{\sqrt{1\mp \frac{ K |V_{fs}| }{\pi V_s}}},
\end{equation}
so that $K_+ >2$ and $K_-<2$.
 Ignoring weak change of $V_s$, one finds
\begin{align}
\label{eq:H+2}
S_+ &= \int_0^\beta \int dx \lbrace \frac{1}{2\pi K_+}[V_{s}^{-1}(\nabla_\tau \theta_{+})^2+ V_{s}(\nabla_x \theta_+)^2]-V_+\cos(2\theta_+ +4\pi \nu x)\rbrace,\\
S_{-} &= \int_0^\beta \int dx \lbrace \frac{1}{2\pi K_-}[V_{s}^{-1}(\nabla_\tau \theta_-)^2+ V_{s}(\nabla_x \theta_-)^2]-V_-\cos(2\theta_-)\rbrace\, ,
\end{align}
where $V_+ \sim V_- \sim V_d$ are the backscattering terms taken for $m=\pm 1$. For $\nu$ different from $1/2$, the term $V_+$ can be dropped so that the first equation describes PSF which is a Luttinger liquid of pairs. In other words, there is algebraic order in the field  $\Psi = \psi_1 \psi_2$. 

The situation with the channel $S_-$ is completely different. Since the critical value of $K_-$ is 2,  the backscattering term becomes relevant so that the fluctuations of the angle $\theta_-$ are suppressed. This implies that the superfluidity in the counter-flow channel is destroyed because density and phase are conjugate variables. Thus, suppression of  fluctuations in $\theta_-$ immediately implies strong fluctuations in the relative phase $\phi_-=\phi_1-\phi_2$ so that the individual  fields $\psi_1\sim \exp(i\phi_1)$ and $\psi_2\sim \exp(i\phi_2)$ lose their coherence.

The corresponding renormalization group (RG) equations (ignoring weak renormalization of the speed of sound \cite{Mathey}) are: 
\begin{align}
\label{eq:RGm1}
 \frac{dg^{-1}}{d\ln l}&=g u^2\\
 \label{eq:RGm2}
\frac{du}{d\ln l}&=2(1-g)u,
\end{align}
where $g=K_-/2$ and $u\propto V_- $ with the proportionality coefficient determined by the small-distance cutoff. 
These equations are the standard Kosterlitz-Thoulless~\cite{BKT} RG equations, where $l$ sets a typical space-time scale (in units $V_s=1$).
At small scales, the flow of $K_-(l)$ starts at the initial value
set by the forward scattering, Eq.~(\ref{eq:Kpm}). Since $K_-(0) <2$, Eq.~\eqref{eq:RGm2}
indicates that $V_-$ becomes more and more relevant as $l$ increases. Simultaneously, the effective Luttinger parameter
$g(l)$ flows to zero as $l\to \infty$.  

The comparison of the prediction of Eqs.~\eqref{eq:RGm1} and~\eqref{eq:RGm2} with {\it ab initio} simulations can be done for $g(l)$. As discussed in the next section, the worm algorithm has a direct access to $g(l)$ where $l$ is determined by the ratio of the system size $L$ to some microscopic scale $L_0$, that is, $l \propto L/L_0$,  provided that the extension along imaginary time $\hbar /T = L/V_s$ establishes the space-time symmetry. 
We will conclude this section by presenting the solution of the RG equations.

A general solution of the system of equations \eqref{eq:RGm1} and \eqref{eq:RGm2} can be expressed in terms of two constants of integration,  $\eta$, $l_0>0$, determined by the initial values of $u$ and $g$, which in their turn are set by the microscopic model \eqref{eq:H}.
If $\eta$ is real, the solution has a form
\begin{align}
\label{eq:sol}
u^2&= 2[ \xi^2 -\eta^2],\\
4\ln\left(\frac{l}{l_0}\right)&\equiv \ln(\xi^2 (l) - \eta^2 ) + \frac{1}{|\eta| }\ln\left(\frac{\xi (l) - |\eta|}{\xi (l) + |\eta|}\right),
\nonumber
\end{align}
where $\xi = \frac{1}{g}-1= \frac{2}{K_-}-1$, $|\xi| > |\eta|$ and $\xi > -1$. 

On the other hand, if $\eta= i |\eta|$, the solution becomes 
\begin{align}
\label{eq:sol2}
u^2&=2[ \xi^2 +|\eta|^2]\\
4\ln\left(\frac{l}{l_0}\right)&\equiv\ln(\xi^2 (l) + |\eta|^2 ) - \frac{2}{|\eta| }\tan^{-1}\left(\frac{|\eta|}{\xi}\right).
\nonumber
\end{align}
This solution is valid if $\xi > -1$.

As will be shown in the section~\ref{sec:QMCN2} the class of solutions pertaining to our simulations data belongs to the separatrix, that is, $\eta\to 0$ and $u=\sqrt{2}|\xi|$ where 
\begin{equation}
\label{sep}
 \ln \vert\xi\vert -\frac{1}{\xi}=2\ln\left(\frac{l}{l_0}\right),\quad \xi = \frac{2}{K_-}-1.
\end{equation}
This equation will be used for the analysis of the QMC simulations where $K_-$ will be determined directly in a progression of system sizes for different values of the interaction.

\subsection{Quantum Monte Carlo Results}\label{sec:QMCN2}
 For simplicity, we restrict the interactions to nearest-neighbor inter-layer attraction.
This corresponds to small filling factor, $\nu <<1$, when the dipole interactions along the tubes can be ignored.

 Our QMC simulations are focused on finding statistics of windings of particle world-lines. The system satisfies periodic boundary condition along space and along imaginary time. Then, these windings are well defined integers $W_x(z)$, $z=1,2$, along space and $W_\tau (z)$, $z=1,2$, along time. As  shown in Ref.\cite{Ceperley}, the mean square fluctuations of $W_x$ and $W_\tau$ determine the superfluid stiffness and superfluid compressibility, respectively.   

The Luttinger parameter $K_-$ can be expressed in terms of $W_x$ and $W_\tau$ as follows (see \cite{Safavi-Naini}):
\begin{equation}
\label{K-W}
K_-=\pi\sqrt{\langle (W_x(0) - W_x(1))^2 \rangle \langle (W_\tau (0) - W_\tau (1))^2\rangle}\; .
\end{equation}

Within the simulations, the RG scale $l$ is set by the system size $L$.
 Practically, we have kept $L \propto \beta$ so that $\langle (W_x(0) - W_x(1))^2 \rangle =\langle (W_\tau (0) - W_\tau (1))^2\rangle$, in order to ensure space-time symmetry. This guarantees that the system remains in its ground state for a given size. 
\begin{figure}[h]
\begin{center}
\includegraphics[trim={-2cm 1cm 0cm 4cm}, clip,width=1\textwidth]{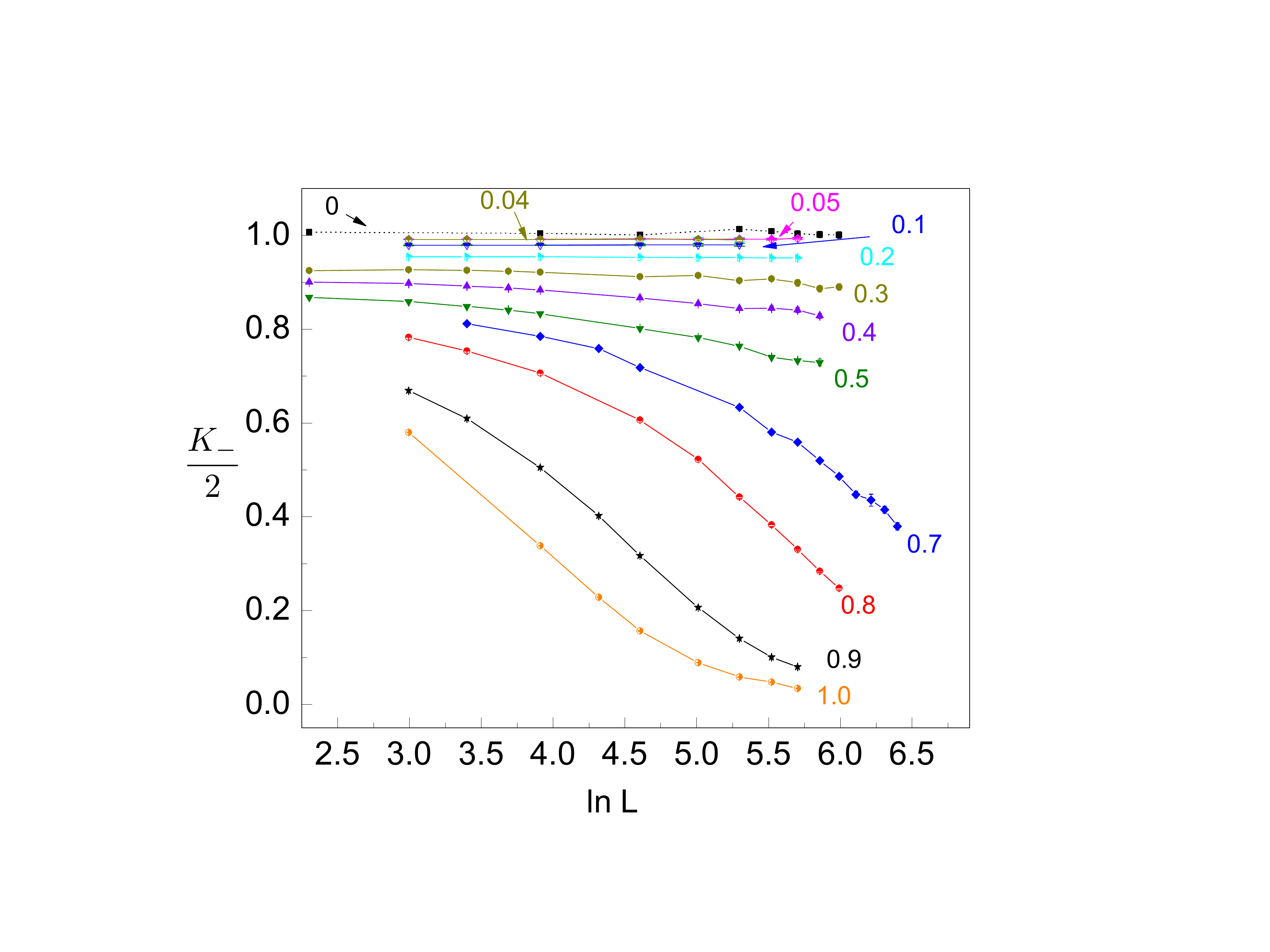}
\caption{(Color online) Numerical data for $K_-/2$ as a function of system size $L$ for different values of the inter-layer interaction $V_d/J$, and in the absence of intra-layer repulsion. The lines are labeled by the corresponding value of $V_d/J$. }
\label{fig:K-N2}
\end{center}
\end{figure}
\begin{figure}[h]
\begin{center}
\includegraphics[trim={-2cm 2cm 2cm 4cm}, clip, width=1\textwidth]{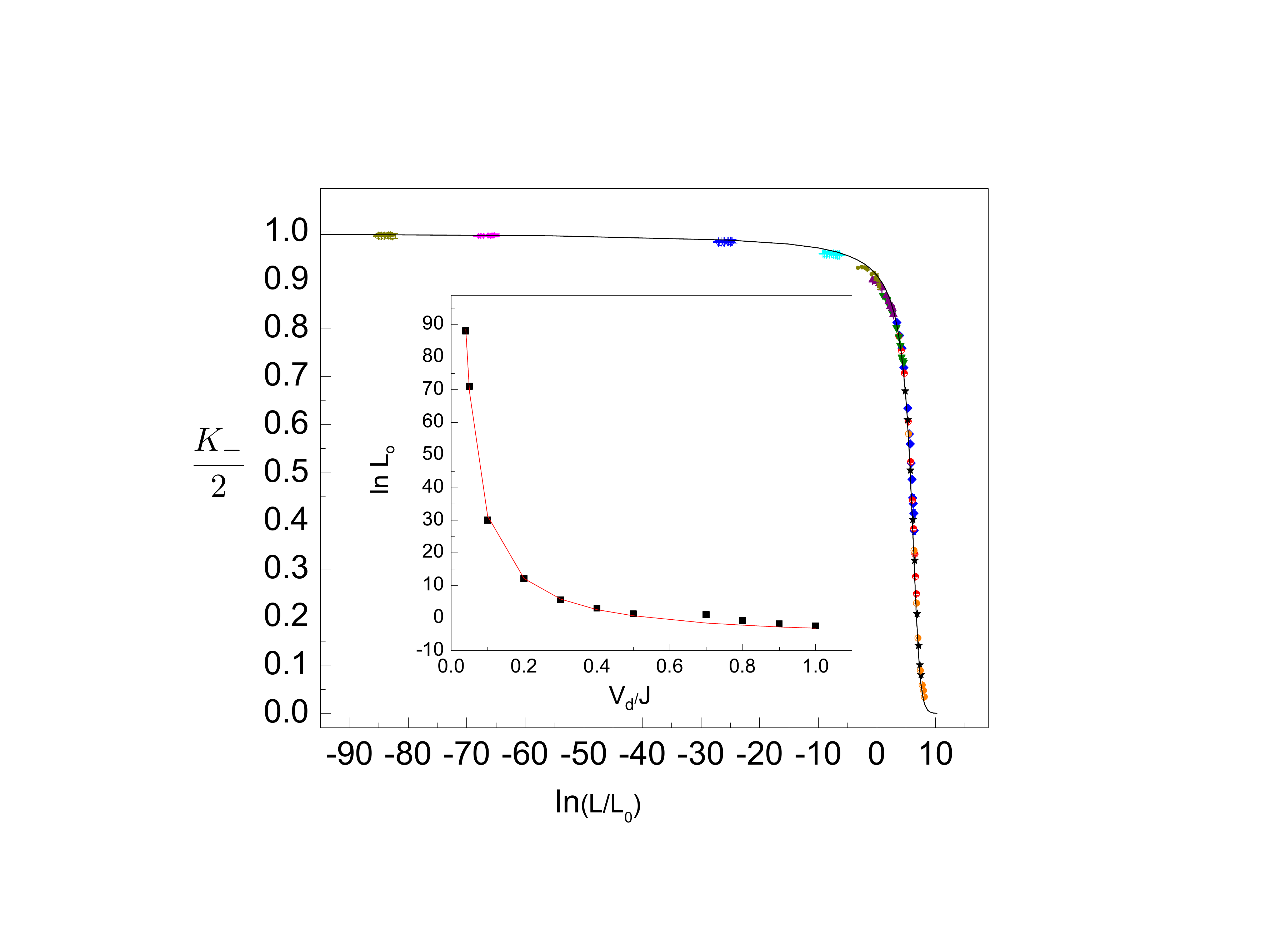}
\caption{(Color online) The data from Fig.~\ref{fig:K-N2} is shown vs $\ln(L/L_0(V_d/J))$ with $L_0(V_d/J)$ chosen in such a way as to achieve the best collapse on the separtrix solution. The size of the symbols is determined by the statistical error bars. The solid line is the solution for the separatrix, Eq.~(\ref{sep}). Inset: plot of the rescaling parameter $\ln L_0$ vs  $V_d/J$. Solid line is the fit by $\ln L_0 = a/(V_d/J)  - b$, with  $ a=3.82, b = 6.96$.  }
\label{fig:master}
\end{center}
\end{figure}

Fig.~\ref{fig:K-N2} presents the raw data for $K_-/2$ obtained from simulations performed at various system sizes. These results allow us to identify the class of solutions to the RG flow best describing the data, and to determine the integration constant $l_0$. We find that within the statistical errors of the simulations, the curves of $K_-/2$ vs $L/L_0$ for various $0< V_d/J < 1$ collapse on to one master curve, shown in figure~\ref{fig:master} which is well described by the separatrix solution given by Eq.~(\ref{sep}).
The master curve has just one fit parameter $l_0$. Thus, for a given value of the interaction $V_d/J$, we have found a unique $L_0=L_0(V_d/J)$ so that the set of the data $K_-(L)$ (for a given $V_d/J$) from Fig.~\ref{fig:K-N2} can be best superimposed on the separatrix solution where $l/l_0=L/L_0$. The result of this procedure for 11 values of $V_d/J$ (with the total of $\sim$100 data points) is presented in Fig.~\ref{fig:master} where the solid line is the separatrix (\ref{sep}). 

The inset of figure~\ref{fig:master} shows a plot of $L_0$ as a function of $V_d/J$. While $L_0$ is a rescaling parameter, it can also be interpreted as the size of a bound dimer. As can be seen from the inset of Fig.~\ref{fig:master}, $L_0$ diverges with decreasing inter-tube interaction, 
\begin{equation}
\label{corM}
L_0 \sim \exp\left(\frac{aJ}{V_d}\right),\quad V_d \to 0,
\end{equation}
 where $a$ is a constant ( $a=3.82$). This divergence proves that the critical value for the formation of the dimer superfluid is $V_d=0$. The excellent agreement between the numerical data and the RG solution (\ref{sep}) over almost 50 orders of magnitude of the (effective) distances as well as the dependence (\ref{corM}) proves that the PSF phase is stabilized for infinitesimal attraction between the layers and that the bosonization can describe very accurately the system properties.

At this point a comment about the meaning of the master curve is in order. Using small $V_d$ in a system of finite size $L$ may not result in an experimentally "visible" formation of PSF because $K_-$ is too close to $K_-=2$. However, for larger size $L$,  $K_-$ will "flow" along the master curve (solid line), i.e. the separatrix, toward smaller values. Such an approach -- matching
the numerical solution with the RG flow in order to find the critical point of the Berezinskii-Kosterlitz-Thouless transition has been pioneered in Ref.~\cite{Kolya_2002}.

\section{Results from N-tube bosonization } \label{sec4}
The formalism of the previous section can be extended to the multilayer case. For a thorough discussion of the $N$-tube bosonization we refer readers to ~\cite{Safavi-Naini}. The bosonization analysis for the general case is more involved since it is not possible to rewrite the action in terms of a sum of actions corresponding to decoupled sectors as it was done for $N=2$. Likewise the $N=2$ case, though, the goal is to study the flow of the amplitudes associated to the various harmonics (as defined in Eq.~\ref{eq:rho}). Here, we present a short qualitative discussion. 

{\em{CSF}.} $\;\;\;$
Similarly to the $N=2$ case, the CSF phase occurs for arbitrarily small inter-tube attraction $V_d$
as long as the filling factors in the tubes are the same.
Due to the spatially non-local nature of dipolar interactions, the CSF phase ~\ref{CS} can form between tubes with the same filling factors regardless of their positions. For example, in a system of $N=6$ tubes where $\nu_1= \nu_2=\nu_5=\nu$ (here we consider $\nu \neq 1/2$) with all other values $\nu_3\neq  \nu_4 \neq  \nu_6\neq \nu$, a CSF will form between tubes $z=1,\,2,\, 5$ while the remaining tubes will form 3 independent SF. Consequently, this system features algebraic orders of individual fields $\psi_3(x), \psi_4(x), \psi_6(x)$, as well as of the CSF field $\Psi_{125}(x) =\psi_1(x) \psi_2(x)\psi_5(x)$, with the individual fields $\psi_1(x)$, $ \psi_2(x)$, $\psi_5(x)$ being disordered.

The important aspect of the solutions found in Ref.~\cite{Safavi-Naini}  is that, in the presence of intra-tube repulsion $V_0>0$ provided by dipole forces,  matching of the filling factors rather than the sign of the inter-tube interaction can determine the nature of possible phases.  
The CSF phase can also occur due to the inter-tube {\it repulsion} (for dipoles polarized perpendicular to the $xz$-plane) as long as the intra-tube repulsion is dominating, and provided the filling factors of the tubes
forming CSF are the same. Such a peculiarity -- attraction due to repulsive interaction -- is the property of  hard-core bosons in 1D \cite{Safavi-Naini}, as it can be seen from the renormalization of the forward scattering amplitude discussed in Eq.~(\ref{eq:Kpm}).
If both the intra- and inter-tube repulsions exist, this equation becomes
\begin{equation}
\label{eq:Kpm2}
K_\pm=\frac{2K}{\sqrt{1+ \frac{ K(V_0 \pm  V_{fs}) }{\pi V_s}}}.
\end{equation}
Thus, the Luttinger parameter $K_-$ for the channel $\theta_-$ becomes $K_-<2$ if $V_0 > V_{fs}$, where $V_{fs}$ is due to inter-tube repulsion. This, as discussed above, implies the suppression of the superflow in this channel (for equal filling factors). [It should be noted that, despite $K_+$ also being below 2, the superfluidity in this channel is protected by the filling factor being different from 1/2, as it was discussed for $N=2$ layers].

%\smallskip
{\em{SCF}.} $\;\;\;$
The supercounterfluid phases can occur between tubes with filling factors complementary to unity. These phases are thresholdless in $V_d$ as well \cite{Safavi-Naini}. If $V_0>|V_{fs}|$, the SCF occurs for attractive inter-tube interaction as can be seen from Eq.(\ref{eq:Kpm2}), provided the filling factors are complementary and different from 1/2. 

An interesting situation occurs when  there are groups of tubes with complementary fillings. Say, there are two sets of tubes, $M_1>1$ and $M_2>1$ such that in the first one the filling factor in each tube is $\nu \neq 1/2$ and in the second one it is $1-\nu$. Then the RG equations show the formation of a {\it pair} of composite superfluids---one per each group of tubes with equal filling factors--- and that these composite superfluids further bind in the counterflow regime.

{\em{CB insulators}. }$\;\;\;$
In the absence of inter-tube interactions, hard-core bosons can form a CB insulator at filling factor $\nu=1/2$ only if the repulsive interaction is strong enough. The situation changes dramatically in the presence of inter-tube interaction. RG equations show that regardless of the sign of the inter-tube interaction even in the limit $V_d \to 0$, a CB solid can be stabilized.
Accordingly, there is no algebraic order in any channel.

\section{Josephson effect between composite superfluids}
\label{sec6}

One can engineer coherent transport through a link between two different composite superfluid, where the transport is controlled by specific  powers of the order parameters corresponding to the two systems across the link (cf. \cite{Kuklov}). In the following we use examples to illustrate this point.  

Consider a CSF characterized by the field $\Psi_k=\psi_1...\psi_k$ formed out of $k$ components. As discussed in Ref. \cite{Safavi-Naini} the transition from $k$ independent condensates to a single condensed field $\Psi_k$ corresponds to the loss of coherence of each individual field $\psi_i$, while coherence is preserved in $\Psi_k$. To emphasize the number of the components we will call such a CSF as $k$-CSF. Here we describe the Josephson coupling between the $k$-CSF and another phase, for example $n$-CSF with $n\neq k$ or a simple superfluid exhibiting one-particle algebraic order in a field $\psi$. This coupling should preserve the global U(1) invariance, which arises as a consequence of the particle number conservation in the system. This requirement excludes couplings of the type $\sim (\Psi_k^* \Psi_n +c.c.)$ or $\sim (\Psi_k^* \psi +c.c.)$.

We begin by considering $k$ tubes where the region $x<0$ is occupied by $k$-CSF while the region $x>0$ contains $k$-SF, that is, $k$ independent superfluids characterized by one-particle fields $\psi_i,\, i=1,2,...,k$ each exhibiting algebraic order. If there is a weak link at $x=0$,     
the corresponding (minimal) Josephson coupling is given by the product $ \sim \Psi^*_k \prod_{i=1}^k\psi_i + c.c.$.
In terms of the corresponding phases the Josephson energy can be written as 
\begin{equation}
E_J= - J_c \cos(\varphi_k - \sum_{i=1}^k \phi_i), 
\label{genJ}
\end{equation}   
where $\Psi_k= |\Psi_k|\exp(i\varphi_k),\, \psi_j=|\psi_i|\exp(i\phi_j)$.

If the fields $\psi_i$ are all identical, that is their phases are locked in as one and the same phase $\phi$ due to the tunneling between the tubes in the region $x>0$, the above relation takes the form
\begin{equation}
E_J= - J_c \cos(\varphi_k - k\phi).
\label{genJ2}
\end{equation}   
In other words, the coherence is preserved only in the transport of $k$ bosons across the junction, and any transfer of a number of bosons different from $k$ would cost finite energy and, therefore, cannot be coherent. It should be noted that higher-order couplings would allow coherent transport of any integer number of groups of $k$ bosons.  We, however, are focusing on the lowest order coupling. 

Next, consider a junction formed between two composite superfluids $k$-CSF and $n$-CSF with $k >1,\, n>1$ and $n\neq k$. In the following we consider the case where $k$ and $n$ have no non-trivial common divisor. 
The corresponding fields exhibiting algebraic orders can be represented as $\Psi_{k}= \prod_i^k \psi_i $ and $\Psi_{n}= \prod_i^n \psi_i $. In this case, the minimal Josephson effect can exist between the {\it products} of the fields $(\Psi_n)^k$ and $(\Psi_k)^n$ with the corresponding Josephson energy being 
\begin{equation}
E_J= - J_c \cos(n\varphi_k - k \varphi_n), 
\label{genJkn}
\end{equation}   
where $\Psi_k= |\Psi_k|\exp(i\varphi_k),\, \Psi_n= |\Psi_k|\exp(i\varphi_n)$.
This form preserves global U(1) invariance of the system.
In other words, the coherent transport occurs by groups of $nk$ atoms. It is worth mentioning that transferring $n$ atoms from $n$-CSF to $k$-CSF would require a compensation for the difference $k-n\neq 0$ by either adding or subtracting atoms. Clearly, such compensation costs finite energy. Similarly, no transfer of $k$ atoms from $k$-CSF to $n$-CSF is possible without spending energy. In contrast, a transfer of a group of $k$ quasi-molecular complexes each made of $n$ atoms from $n$-CSF to $k$-CSF costs no energy and, therefore, can be coherent. [Conversely, a group of $n$ complexes each built out of $k$ atoms can be coherently transferred from $k$-CSF to $n$-CSF].

If $p>1$ is the greatest common divisor, the minimal coupling can be achieved between the products $(\Psi_n)^{k/p}$ and $(\Psi_k)^{n/p}$ with the Josephson energy
\begin{equation}
E_J= - J_c \cos\left(\frac{n}{p}\varphi_k - \frac{k}{p} \varphi_n\right). 
\label{genJknp}
\end{equation}   
In this case, the coherent transport proceeds by increments of $kn/p$ bosons. More specifically, $k/p$ groups of $n$ bosons are transferred from $n$-CSF to $k$-CSF and $n/p$ groups of $k$ bosons are transferred from $k$-CSF to $n$-CSF.

The most direct implication of the above analysis is an unexpected quantization of the rotational flux. We consider it below for the case of $k$-CSF and $n$-CSF quantum liquids confined to a ring geometry and separated by two weak links. 

\subsection{Ring with  $k$-CSF $n$-CSF junctions and velocity quantization}

In a simple superfluid of bosons each of mass $m$, the  quantization of the velocity circulation  is characterized by the quantum of circulation $\Phi_0=h/m$, where $m$ is the {\it bare} atomic mass. This relation follows from the irrotational nature of superfluidity, the phase $\phi$ of the superfluid being defined modulo $2\pi$, and the requirement of the Galilean invariance. Accordingly, in the $k$-CSF phase (where the coherent transport occurs by groups of $k$ bosons) the circulation quantum is given by $\Phi_k=h/(mk)$ and $ \Phi_n=h/(mn)$ in the $n$-CSF phase.  
In the following we address how the quantization changes if there are two composite superfluids confined in the ring geometry and connected by weak links using two different approaches. 

\begin{figure}[h]
\begin{center}
\includegraphics[trim={4cm 11cm 13cm 2cm}, clip, width=0.8\textwidth]{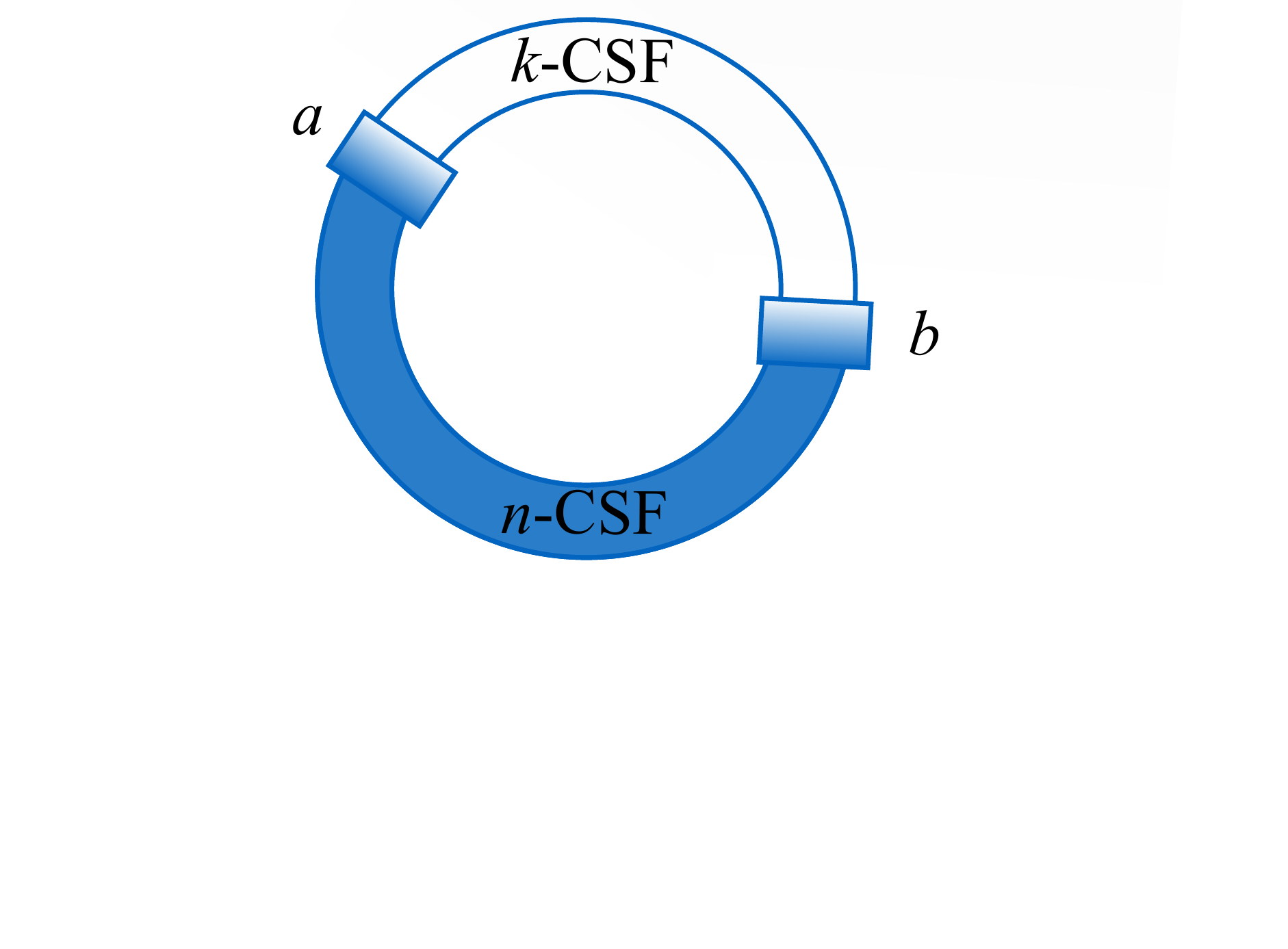}
\vspace{-3cm}
\caption{(Color online) Schematics of the ring: $k$-CSF phase as an open area along the ring, and  $n$-CSF phase -- as the solid one. Weak links $a$ and $b$ are indicated  by rectangles. }
\label{fig:loop}
\end{center}
\end{figure}

Consider a ring, Fig.~\ref{fig:loop}, of radius $r$ containing two composite superfluids $k$-CSF and $n$-CSF and rotating at angular velocity $\Omega$ around the axis through the ring center and perpendicular to the ring plane.
The Josephson energy $E_J$ of the system as a function of the difference of the phases $\varphi_k$ and $\varphi_n$  is the sum of the energies written for
two (identical)  junctions
\begin{equation}
E_J= - J_c \left[\cos\left(\frac{k}{p}\varphi_n(a) - \frac{n}{p}\varphi_k(a)\right) + \cos\left(\frac{k}{p}\varphi_n(b) - \frac{n}{p}\varphi_k(b)\right)\right].  
\label{genJ5}
\end{equation}   

Without loss of generality we can set $\varphi_k(a)=0$ and $\varphi_n(b)=0$. Then, integrating the velocity along the ring, we find $I_n=\hbar \varphi_n(a)/(mn)$ along the $n$-CSF arc and $I_k=\hbar \varphi_k(b)/(mk)$ along the $k$-CSF arc. 
If the arc-lengths $L$ are large, the energy of the currents along the rim is $\propto 1/L$ and can be ignored and one should only consider the Josephson energy given by Eq.~(\ref{genJ5}). This energy is periodic in the phases $\varphi_k(b)$ and $\varphi_n(a)$  
with the corresponding periods $2p\pi/n$ and $2p\pi/k$. Such a periodicity is the analog to the phase $\phi$ of a simple superfluid being defined modulo $2\pi$. Thus, the quantum of the circulation along the ring can be written as $\Phi_{nk}=I_k + I_n=hp/(mnk)=\Phi_0 p/(nk)$.  This result implies that the coherent current circulating along the ring can be induced at special values of the angular frequency with the period 
\begin{equation}
\delta \Omega =\frac{p}{nk} \frac{\hbar}{m r^2}  . 
\label{dO}
\end{equation}
Here the $nk/p$ is the smallest number of atoms which can pass through the junctions in the coherent manner. So, the situation resembles the case of the $(nk/p)$- CSF occupying the whole ring.

One can also address this problem from the perspective of Galilean invariance.
The velocity $v$ of the flow in the rotating frame becomes $v -v_\Omega$, where $v= \hbar \nabla \phi /m$ and $v_\Omega = \Omega r$. For a thin ring the gradient symbol reduces to a derivative along the rim. 
The rim velocity plays the role of the gauge field $A= m v_\Omega /\hbar$ for the phase $\phi$. In other words, $\nabla \phi \to \nabla \phi -A$ . In the $k$-CSF and $n$-CSF phases, the respective coherent gradients become $\nabla \varphi_k - k A$ and $\nabla \varphi_n - n A$. 
These gauge invariant gradients determine superfluid (coherent) transport in $k$-CSF and $n$-CSF, respectively. 
As mentioned above, in the presence of a junction the coherence is preserved for groups of $kn/p$ bosons. Thus, the current 
 retaining coherence in both phases can be represented by the following gauge invariant gradients $J=(n/p)\nabla \varphi_k - (kn/p) A$ in the $k$-CSF and  $J= (k/p)\nabla \varphi_n - (kn/p) A$ in the $n$-CSF.
Accordingly, the circulation quantization condition becomes
\begin{equation}
\oint J dl = 2\pi f, \quad f=0,\pm 1, \pm 2,...
\label{quant}
\end{equation}
For two junctions, the above expression can be used to obtain
\begin{equation}
 \frac{n}{p} \varphi_k(a) - \frac{k}{p} \varphi_n(a) + \frac{k}{p} \varphi_n(b) - \frac{n}{p} \varphi_k(b)=2\pi f + \frac{nk}{p} \oint A dl.   
\label{quant2}
\end{equation}
Using this expression  in Eq. (\ref{genJ5}) we obtain
\begin{equation}
E_J= -  2 J_c \cos\left(\frac{nk}{2p} \oint A dl\right) \cos\left( \frac{nk}{2p}\oint A dl + \frac{n}{p} \varphi_k(b) - \frac{k}{p}\varphi_n(b)\right). 
\label{genJ55}
\end{equation}
Thus, the Josephson coupling becomes modulated $\sim  J_c \cos\left(\frac{nk}{2p} \oint A dl\right)$ so that it vanishes at 
specific values of $\Omega$ with the period $\delta \Omega$ determined by $\frac{nk}{p} \oint A =2\pi$, which gives Eq.(\ref{dO}).

It is important to notice the factor $p/(nk)$, reduces the flux period when compared to the standard situation, that is, two junctions in a ring containing simple superfluids. In general, lowering the period leads to an increased sensitivity to rotations.
A similar situation occurs in the case when $n=k$, when the maximum divisor is $p=n=k$ and the factor $p/(k n)$ becomes simply $1/n$. Using two composite superfluids with, for example, $k=n+1$ results in the reduction of the period by the factor $1/n(n+1)$.
   
\section{Summary and outlook}
\label{sec7}
In summary, we considered a system of dipolar bosons trapped in a stack of one-dimensional optical lattice layers (tubes) with all tubes lying on the same plane. The direction of polarization is perpendicular to the tubes and can lie in the plane or perpendicular to it. Within this geometry, in the first case the dipolar interactions between bosons are attractive  in different tubes and repulsive within each  tube. In the second case, the interactions are purely repulsive.
 Based on previous results~\cite{Safavi-Naini}, we provided a description of the various quantum phases stabilized in this system depending on interactions and the filling factor in each layer. These phases include independent superfluids, composite superfluids, composite solids and supercounterfluids. 
Composite superfluids and supercounterfluids are found to be threshold-less with respect to the dipolar interaction strength as derived by a combination of bosonization and renormalization group techniques. We described in detail the bosonization formalism for the case of two tubes and verified its main prediction -- the absence of the threshold for the formation of paired superfluid --  by {\it ab initio} QMC simulations. We have also discussed how bosonization predicts that the result obtained for two tubes also holds for arbitrary number of tubes.

Moreover, we proposed how the phases stabilized in this multilayer geometry can be used to create generalized Josephson junctions, with applications to higher sensitivity rotation sensors. In particular, utilizing junctions between composite superfluids made of $k$ and $n$ components trapped in ring-shaped optical lattice layers, one can engineer a sensor for rotations analogous to a SQUID. In this setup, the rotational flux quantum is significantly reduced, by a factor $p/(k n)$, where $p$ is the maximal common divisor. Hence, the accuracy of this rotation sensor can be significantly higher than in the cases of simple superfluids or identical composite superfluids. 

In this work we set out to elucidate the multi-faceted potential of the composite phases of dipolar bosons with respect to fundamental questions and possible applications in various devices. One of the most interesting features is the non-trivial quantization of circulation of superfluid velocity. We have explored this question for the simplest geometry and we don't exclude that more interesting features can be uncovered for various networks and combinations of the composite phases. Moreover, questions associated with the possibility of practical realization and detection provide another interesting direction for future research.

\section*{Aknowledgments} 
This work was supported by the NSF grants PHY1552978,   PHY1314469 and MIUR (PRIN 2010LLKJBX). The computation for this project was performed at the OU Supercomputing Center for Education and Research (OSCER) at the University of Oklahoma (OU) as well as at the  CUNY HPCC  supported by  NSF Grants CNS-0855217, CNS-0958379 and ACI-1126113.
%%%%%%%%%%%%%%%%%%%%%%%%%%%%%%%%%%%%%%%%%%%%%%%%%%%%%%%%%%%%%%%%%%%%%%%%%%%%%%%%%%%%%%%%%%%%

\section*{References}

\bibliographystyle{iopart-num}

\bibliography{atomtronics}

\end{document}